# Doing good vs. avoiding bad in prosocial choice: A refined test and extension of the morality preference hypothesis

Ben M. Tappin [1*] and Valerio Capraro [2]

[1]Royal Holloway, University of London, [2]Middlesex University London

*Corresponding author: benmtappin@googlemail.com

## Abstract

Prosociality is fundamental to human social life, and, accordingly, much research has attempted to explain human prosocial behavior. Capraro and Rand (*Judgment and Decision Making, 13,* 99-111, 2018) recently provided experimental evidence that prosociality in anonymous, one-shot interactions (such as Prisoner's Dilemma and Dictator Game experiments) is not driven by outcome-based social preferences – as classically assumed – but by a generalized morality preference for "doing the right thing". Here we argue that the key experiments reported in Capraro and Rand (2018) comprise prominent methodological confounds and open questions that bear on influential psychological theory. Specifically, their design confounds: (i) preferences for efficiency with self-interest; and (ii) preferences for action with preferences for morality. Furthermore, their design fails to dissociate the preference to do "good" from the preference to avoid doing "bad". We thus designed and conducted a preregistered, refined and extended test of the morality preference hypothesis (N=801). Consistent with this hypothesis, our findings indicate that prosociality in the anonymous, one-shot Dictator Game is driven by preferences for doing the morally right thing. Inconsistent with influential psychological theory, however, our results suggest the preference to do "good" was *as* potent as the preference to avoid doing "bad" in this case.

***Author note:*** BMT gratefully acknowledges funding from the Economic and Social Research Council. We thank Patrick Heck for comments on an earlier draft.

**Introduction**

People often pay costs to benefit others; they behave *prosocially*. Fundamental to human social life (Fehr & Gächter, 2002; Gintis et al, 2003; Nowak, 2006; Tomasello, 2014), prosocial behavior is often explained by appeal to reciprocity. If I pay a cost to help you today, you – or others who learn about my behavior – are more likely to help me tomorrow (Nowak & Sigmund, 2005; Rand & Nowak, 2013; Trivers, 1971). Defying explanations of this kind, however, prosocial behavior is frequently observed in contexts where opportunities for reciprocity are absent. For example, in anonymous, one-shot interactions, individuals often forego some amount of self-interest to the benefit of strangers (Camerer, 2003).

Behavioral economists have classically sought to explain such behavior by assuming that individuals have preferences for minimizing inequity or maximizing efficiency (i.e., social welfare) (Bolton & Ockenfels, 2000; Capraro, 2013; Charness & Rabin, 2002; Engelmann & Strobel, 2004; Fehr & Schmidt, 1999; Levin, 1998). According to these influential frameworks, prosocial individuals derive utility – psychological benefit – from particular social *outcomes*; thus, realizing those outcomes offsets the cost of behaving prosocially.

A recent alternative perspective is that individuals derive utility from performing *actions* they perceive to be morally right (Bicchieri, 2005; DellaVigna et al., 2012; Huck et al., 2012; Krupka & Weber, 2013). This perspective accords with evidence from social psychology that individuals derive utility from seeing themselves in a positive moral light (Aquino & Reed, 2002; Dunning, 2007) and, in addition, that prosocial individuals in particular view opportunities for prosocial action in moral terms; for example, by considering what the morally "right" action is (Liebrand et al., 1986; Weber et al., 2004).

Building on these converging lines of evidence, recent experimental work advanced the hypothesis that a *generalized morality preference* – rather than preferences for minimizing inequity or maximizing efficiency per se – drives prosocial behavior in anonymous, one-shot interactions (Capraro & Rand, 2018). In other words, that a simple preference for doing (what is perceived to be) the morally "right" thing underpins individuals' prosociality in these contexts.

In their key experiments, Capraro and Rand (2018) used a "Trade-Off Game" (TOG) to empirically dissociate the hypothesized morality preference from outcome-based social preferences for equity and efficiency. In the TOG, participants made a unilateral choice about how to allocate money between themselves and two other (passive) people. While one choice minimized inequity – all participants earned the same amount – the other choice maximized efficiency – participants earned different amounts, but, together, the group earned more. This design effectively pitted preferences for equity and efficiency against one another; creating a decision context where the morally "right" choice was ambiguous. The researchers found that, framing *either* choice as the morally appropriate one dramatically affected participants' choices, such that the majority chose the option framed as morally appropriate; be that the equitable *or* efficient choice.

To support the inference that these moral considerations drive prosociality, however, required additional evidence. To that end, participants also completed, in addition to the TOG, a canonical prosocial choice task; either the Dictator Game (DG), or the Prisoner's Dilemma (PD). In the latter tasks, participants made a unilateral choice about how much money to donate to a new (passive) person (DG), or a simultaneous bilateral choice whether to cooperate with a new person (PD), respectively.

The key finding in Capraro and Rand (2018) was that participants who made the choice framed as morally appropriate in the TOG – be that the equitable choice *or* the efficient choice – were consistently more prosocial in the DG and PD; donating and cooperating (respectively) more than participants who chose otherwise in the TOG. Crucially, this result is *inconsistent* with stable outcome-based preferences for equity or efficiency as explanations for prosociality, which do not predict an association between moral framing in the TOG and prosociality in a different task, such as the DG/PD. The result is instead consistent with the morality preference hypothesis, which predicts that individuals sensitive to which choice is morally right in the TOG – as revealed by the moral framing of those choices – are also revealed to be more prosocial in the DG/PD; where, in contrast to the TOG, the morally right choice is *unambiguous* (Krueger & Acevedo, 2007; Krueger & DiDonato, 2010).

The implication of Capraro and Rand's (2018) findings is important: They suggest their data renders the classic approach to understanding prosocial choice through social preferences insufficient and, in particular, that an account based on a fluid preference for "doing the morally right thing" is superior. However, their key evidence derives from an experimental design that contains several prominent methodological confounds, and leaves open important questions regarding the mechanism of the hypothesized morality preference. Below we expand on these issues.

**Self-interest**

Consider the choice outcomes in the TOG. The *equitable* choice always provided the participants – the chooser, and two passive recipients – the same allocation; 13 Monetary Units (MU) each. The *efficient* choice, in contrast, always provided the chooser with 15 MU, and the passive recipients 23 MU and 13 MU, respectively. Thus, while the efficient choice clearly results in greater overall gains for the group – at the cost of equity, as intended – it also results in *greater gains for the chooser themselves*. In other words, the choice option meant to reveal a preference for efficiency is confounded with self-interest. A plausible consequence of this confound is an *overestimate* of the proportion of individuals with a preference for efficiency. An overestimation of this kind may have affected the key result – an association between TOG choice and prosociality in the DG/PD – in two ways.

First, it may have *inflated* the association between TOG choice under the *equitable*-is-moral frame, and prosociality in the DG/PD. Specifically, this association may not have been driven by participants with a genuine morality preference – who choose the equitable option under this TOG frame, and the prosocial option in the DG/PD – but, rather, by self-interested participants – who choose the *efficient* option under this TOG frame, and the *self-interested* option in the DG/PD. Indeed, in the worst case, the behavior of self-interested participants could *fully account* for the observed association between TOG choice under the equitable-is-moral frame, and prosociality in the DG/PD.

Second, by the opposite logic, the overestimation of individuals with a preference for efficiency may have *deflated* the association between TOG choice under the *efficient*-is-moral frame, and prosociality in the DG/PD. This is because some participants making the efficient choice under that TOG frame did so *not* because of a general morality preference nudged by the framing, but, rather, for their own self-interest. Crucially, these participants would *not* have chosen prosocially in the DG/PD, thereby deflating the observed association between the two choices.

These issues directly affect the key evidence – an association between TOG choice and prosociality in the DG/PD – supporting the morality preference hypothesis. A remedy to these issues is to remove self-interest from the equation by design.

**Action-inaction asymmetry**

Not only do the *efficient*-is-moral and *equitable*-is-moral frames differ in the labels used to describe the two choice options, but, in addition, they differ in which is the *active* choice and which is the *passive* choice. Specifically, in the *efficient*-is-moral frame, participants start with an equitable allocation (13 MU each), while in the *equitable*-is-moral frame they start with an efficient allocation (15, 23, and 13 MU, respectively). In other words, the moral choice is always framed as an *active* choice to change these initial allocations. Choice frame is thus confounded with active/passive frame.

A substantial body of work in social, moral, and decision-making psychology indicates that humans perceive *inaction* differently than *action* (Baron & Ritov, 2004; Spranca et al., 1991). For example, regret is greater for actions that lead to negative outcomes than for *inactions* that lead to the same negative outcomes (Feldman & Albarracín, 2017; Zeelenberg et al., 2002); individuals are biased towards maintaining the status quo in decision-making (Samuelson & Zeckhauser, 1988); and, in moral judgment, harms caused by *action* are considered worse than the same harms caused by *inaction* (Cushman et al., 2006). Finally, most relevant here, action framing influences engagement in prosocial behavior (Teper & Inzlicht, 2011), and there is considerable variation in *who* exhibits action-inaction asymmetries (Baron & Ritov, 2004).

Given this evidence, it is probable that the confounding of choice frame with active/passive frame over- or under-estimated the proportion (and types) of individuals choosing the morally-framed option in the TOG; with unknown consequences for the key association between TOG choice and prosociality in the DG/PD. Decoupling these frames is necessary to make clear inferences about the effect of choice frame in the TOG.

**Doing good vs. avoiding bad**

An influential hypothesis in social psychology is that immoral, negative, or otherwise "bad" stimuli weigh more heavily than their "good" counterparts in human cognition and behavior (Baumeister et al., 2001; Rozin & Royzman, 2001; Vaish et al., 2008; see Corns, 2018 for a recent critique).

Consistent with this hypothesis, recent evidence suggests that "self-righteousness" – manifested in, for example, the *average person* rating themselves morally superior to the average person (Tappin & McKay, 2017) – are greater for immoral than moral stimuli (Klein & Epley, 2016, 2017). Relatedly, the correlation between individuals' life satisfaction and their self-perception is reportedly stronger if the latter is computed as the distance between individuals' "real" and "undesired" selves vs. between their "real" and *"desired"* selves (Ogilvie, 1987). In other words, those data suggest the type of person individuals want to *avoid* being weighs more heavily (in their life appraisal) than the type of person they would ideally like to be. A similar asymmetry manifests in the psychology of moral regulation. In particular, in the distinction between *pro*scriptive morality – what we *should* do and be – and *pre*scriptive morality – what we should *avoid* doing and being. Whereas the former is considered discretionary and a matter of personal preference, the latter is considered mandatory and strict (Janoff-Bulman et al., 2009).

To implement the choice framing in the TOG, the two choice options were *jointly* framed as moral and immoral, respectively. Individuals choosing the morally-framed option may thus have been motivated by a preference to do "good" (e.g., a desire to be moral), or motivated by a preference to avoid "bad" (e.g., an aversion to being immoral). These distinct preferences are confounded in the TOG design. Given the preceding evidence, it is plausible that individuals' choices were motivated more by a preference to avoid "bad" than to do "good". Furthermore, assuming this hypothesis, a further plausible hypothesis is that participants who were motivated by a preference to do "good" (vs. avoid "bad") in the TOG were more likely to behave prosocially in the DG/PD. For example, because the preference to avoid bad may reflect a general desire to avoid punishment, whereas a preference to do good may reflect a desire to do good for its own sake. That is, the latter preference is more diagnostic of true prosocial motivation.

**The current study**

Here we address the methodological confounds and open theoretical questions in Capraro and Rand (2018), and, thus, provide a refined and extended test of the morality preference hypothesis. To do so, we design and implement an improved Trade-Off Game (TOG), and test for an effect of choice frame on TOG choice (Hypothesis 1), and for an association between framing of the TOG and prosociality in a different task, the DG (Hypothesis 3). We also test two novel hypotheses bearing on existing psychological theory. First, that the effect of choice frame on TOG choice is greater under an avoid "bad" than do "good" moral frame (Hypothesis 2) (cf. Baumeister et al., 2001; Rozin & Royzman, 2001; Vaish et al., 2008). Second, that the association between choosing the morally-framed option in the TOG, and prosociality in the DG, is greater under a do "good" than avoid "bad" moral frame (Hypothesis 4).

## Methods

The hypotheses, design, sampling and analysis plan were preregistered on the Open Science Framework (protocol: https://osf.io/9nphs/).

**Participants**

We sought to collect N=200 participants per treatment, giving a total N=800. We determined this sample size by multiplying the N-per-treatment in Capraro and Rand (2018) study 3 by 1.5x; the study most conceptually similar to that which we reproduce here. Sensitivity power analyses (reported in SI) for our key hypothesis tests indicated we had sufficient power (>.80) to detect standardized effect sizes conventionally considered small ($r$ =.10). A total of N=801 participants completed the study. Participants were recruited online via Amazon's Mechanical Turk (AMT) (for the validity of AMT, see e.g., Arechar et al., 2018; Horton et al., 2011; Paolacci et al., 2014; Thomas & Clifford, 2017), and were located in the US at the time of taking part. All participants provided informed consent. This study was reviewed and approved by Middlesex University, and Royal Holloway, University of London ethics procedures.

**Procedure**

Participants began by playing a Dictator Game (DG). In the DG, they were given $0.10 and they had to decide how much, if any, to give to another anonymous participant who received no starting money allocation. Participants could donate in increments of $0.01; from $0.00 to $0.10. The participant was informed that the other person had no active choice and would only receive what they decide to give. We asked two comprehension questions to ensure that participants understood the payoff structure of the DG prior to their decision. Specifically, we asked which choice (1) maximized their *own* payoff, and which choice (2) equalized their payoff with that of the other person. Participants who failed either or both comprehension questions were prevented from completing the survey (this condition was made explicit in the consent form). Those who passed the comprehension questions were then asked to make their DG decision.

Following the DG, participants played an improved Trade-Off Game (TOG). In this TOG, participants ("choosers") had to decide between two choice options that affected their own payoff and the payoff of two other people; the latter being passive recipients who did not make any choices. One option was "equitable", in the sense that it minimized payoff differences among the three participants; specifically, they each earned $0.13. The other option was "efficient", in the sense that it maximized the sum of the payoffs of the three participants; specifically, the chooser earned $0.13, while the other two people earned $0.23 and $0.13, respectively. Importantly, in this improved TOG design, because the chooser earns $0.13 by making *either* choice, the confounding of self-interest with preferences for efficiency is eliminated. Furthermore, because participants are not told that one or the other state of money distribution ([13, 13, 13] or [13, 23, 13]) initially holds, *both* choice options are rendered equal in terms of active/passive frame.

Before reading the TOG instructions, participants were randomly assigned to one of four versions of the TOG, each corresponding to a particular framing combination in a 2x2 between-subjects design:

- TOG frame: **Give – Do Good**
- TOG frame: **Give – Avoid Bad**
- TOG frame: **Equalize – Do Good**
- TOG frame: **Equalize – Avoid Bad**

We experimentally manipulate whether the efficient ("Give") or equitable ("Equalize") choice is framed as morally appropriate (choice frame[1]: Give, Equalize).

---

[1] Following Capraro & Rand (2018), in our preregistered protocol and analysis script we labelled the efficient-is-moral frame the "Give" frame, and the equitable-is-moral frame the "Equalize" frame. For consistency, we follow that convention here.

Furthermore, we also manipulate whether the moral framing emphasizes doing "good" or avoiding "bad" (moral frame: Do Good, Avoid Bad):

- Under the **Give – Do Good** frame, the efficient option is labelled "be generous, Option 1", and the equitable option is "Option 2"
- Under the **Give – Avoid Bad** frame, the efficient option is labelled "Option 2", and the equitable option is "be ungenerous, Option 1"
- Under the **Equalize – Do Good** frame, the efficient option is labelled "Option 2", and the equitable option is "be fair, Option 1"
- Under the **Equalize – Avoid Bad** frame, the efficient option is labelled "be unfair, Option 1", and the equitable option is "Option 2"

Importantly, notice that the experimental manipulation of moral frame decouples the preference to do "good" from the preference to avoid "bad". After making their decision in the TOG, participants provided standard demographic information, at the end of which they were given the completion code needed to submit the survey on AMT. After the end of the survey, we downloaded the data file and computed the bonuses, which were paid on top of the base participation fee received by all participants ($0.50). No deception was used. We refer to the SI for verbatim experimental instructions (available here: https://osf.io/m7w2s/). We report all measures, manipulations and exclusions.

## Results

Data analysis was conducted in R (v.3.4.0, R Core Team, 2017) using RStudio (v.1.1.423, RStudio Team, 2016). R packages used in analysis and figures: *ggplot2* (v.2.2.1, Wickham, 2009), *plyr* (v.1.8.4, Wickham, 2011), *dplyr* (v.0.7.4, Wickham et al., 2017), *reshape* (v.0.8.7, Wickham, 2007), *gridExtra* (v.2.3, Auguie & Antonov, 2017), *effsize* (v.0.7.1, Torchiano, 2017), *data.table* (v.1.10.4-3, Dowle et al., 2017). The raw data and code to reproduce all results and figures in this paper are available at https://osf.io/x5stj/.

### Data exclusions

N=288 (26.45%) participants answered one or more of the comprehension questions incorrectly, or did not answer these questions, and were thus prevented from completing the study (following our preregistered protocol). Of the remaining N=801 participants who completed the study, there were N=15 (1.87%) duplicate responses according to participants' IP address/unique Mechanical Turk ID. In line with our preregistered protocol, we excluded these duplicates, retaining the earliest responses only—defined by the date/time they began the study. Finally, after these exclusions, we identified N=2 (0.25%) participants that dropped out of the study prior to making their decision in the TOG, and are thus unable to be included in the analysis (leaving N=784 for analysis).

### Hypotheses 1 & 2

*Preregistered analyses*

We first test whether participants were more likely to choose the efficient option in the TOG under the "give" choice frame than under the "equalize" choice frame (Hypothesis 1). We then test whether this framing effect was stronger under the "avoid bad" moral frame than under the "do good" moral frame (Hypothesis 2). To that end, we fit a binomial logistic regression model with two dummy-coded treatment variables as predictors: choice frame [0=equalize frame, 1=give frame] and moral frame [0=avoid bad, 1=do good], and choice in the TOG as the dependent variable [TOG choice: 0=equitable choice, 1=efficient choice].

Consistent with Hypothesis 1, choice frame predicted TOG choice in the expected direction, Odds Ratio (OR) = 3.55, 95% CI [2.34, 5.40], $Z = 5.94$ $p < .001$. A majority of participants chose the efficient option under the give choice frame (69.21%), whereas only a minority of participants chose this option under the equalize choice frame (39.90%).

Inconsistent with Hypothesis 2, there is no statistically significant interaction between choice frame and moral frame, OR = 0.91 [0.50, 1.64], $Z = -0.32$, $p = .752$. In other words, the effect of choice frame appeared largely independent of whether the choice was framed as "doing good" or "avoiding bad". Both H1 and H2 results remain similar after adjusting for age, gender [not female=0, female=1] and education [0=less than college, 1=college or above] in the model: Main effect of choice frame OR = 3.74 [2.45, 5.72], $Z = 6.11$, $p < .001$; interaction between choice frame and moral frame OR = 0.89 [0.49, 1.61], $Z = -0.39$, $p = .697$. The proportion of choices in each of the four treatments is displayed in Figure 1.

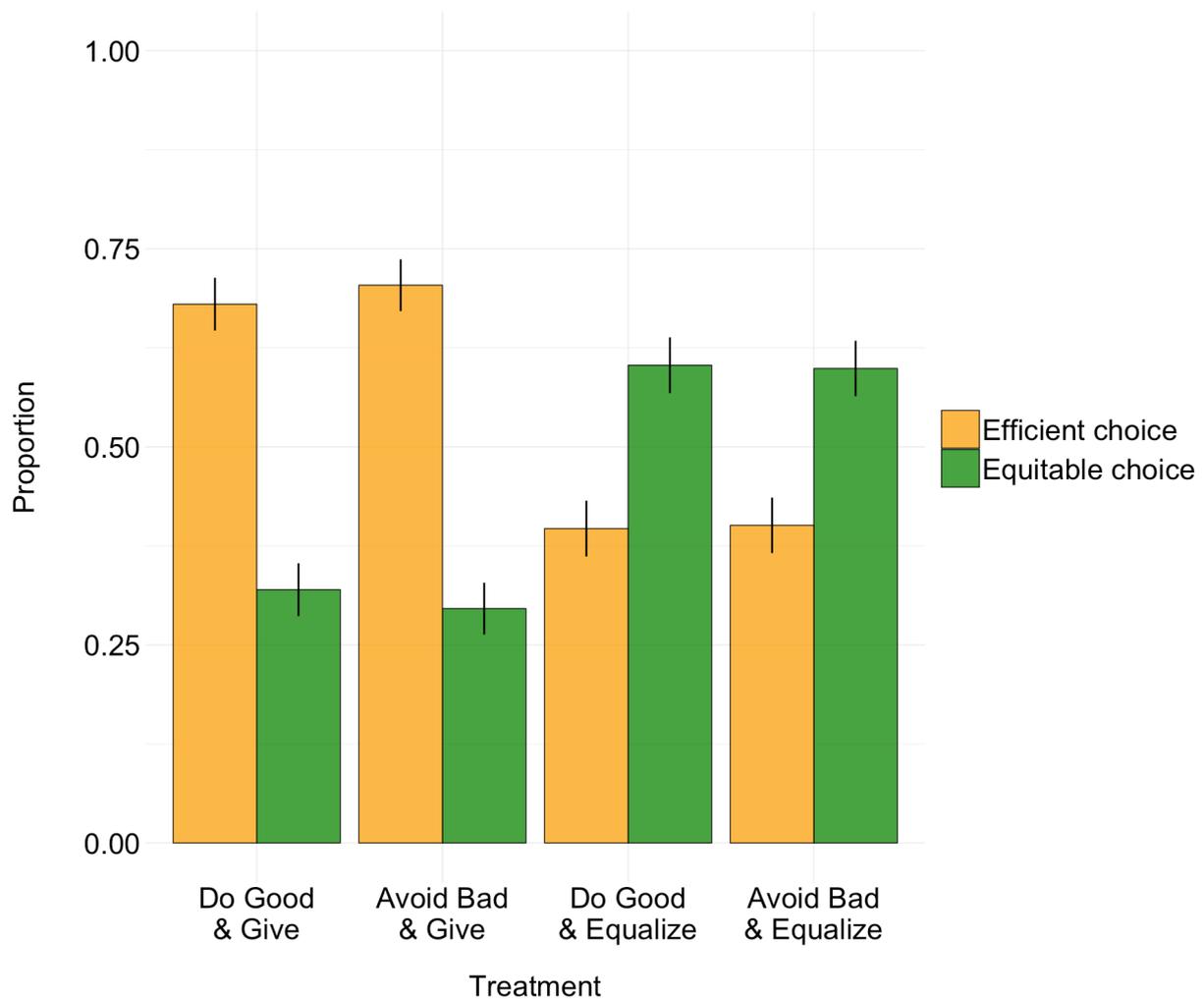

**Figure 1. Proportion of efficient and equitable choices as a function of treatment.** *Error bars are ± 1 SEM.*

**Hypothesis 3**

*Preregistered analyses*

Next, we test the hypothesis that participants who make the 'moral' choice in the TOG – that is, choose the efficient option under the "give" frame, or choose the equitable option under the "equalize" frame – donate more to their partner in the DG (Hypothesis 3). We fit a linear regression model with two dummy-coded variables as predictors: TOG choice frame [0=equalize frame, 1=give frame] and the choice the participant made in the TOG [0=equitable choice, 1=efficient choice], respectively. The DV is amount donated in the DG [from 0 to 10].

Consistent with Hypothesis 3, there is an interaction in the predicted direction, b = 1.68, SE = 0.40, t = 4.17, p <.001. Under the equalize frame, participants who made the equitable choice donated more in the DG (M = 3.40, SD

= 2.60) than participants who made the efficient choice (M = 2.38, SD = 2.57), t (389) = 3.81, p <.001, hedges' *g* = 0.39, 95% CI [0.19, 0.60]. This pattern was reversed under the give frame. There, participants who made the equitable choice donated *less* in the DG (M = 2.53, SD = 2.75) than participants who made the efficient choice (M = 3.19, SD = 2.80), t (391) = -2.19, p =.029, hedges' *g* = -0.24 [-0.45, -0.02]. (Note: All t-tests are post-hoc.) This interaction between choice frame and TOG choice remains similar after adjusting for age, gender, and education, b = 1.67, SE = 0.40, t = 4.14, p <.001. The interaction pattern is displayed in Figure 2 (panel A). As seen in Figure 2, DG donations follow an approximately bimodal distribution peaking over donations of 0 and 5. We thus conducted exploratory analyses to test the robustness of the preceding linear regression results (reported in SI). These were consistent with the linear regression models.

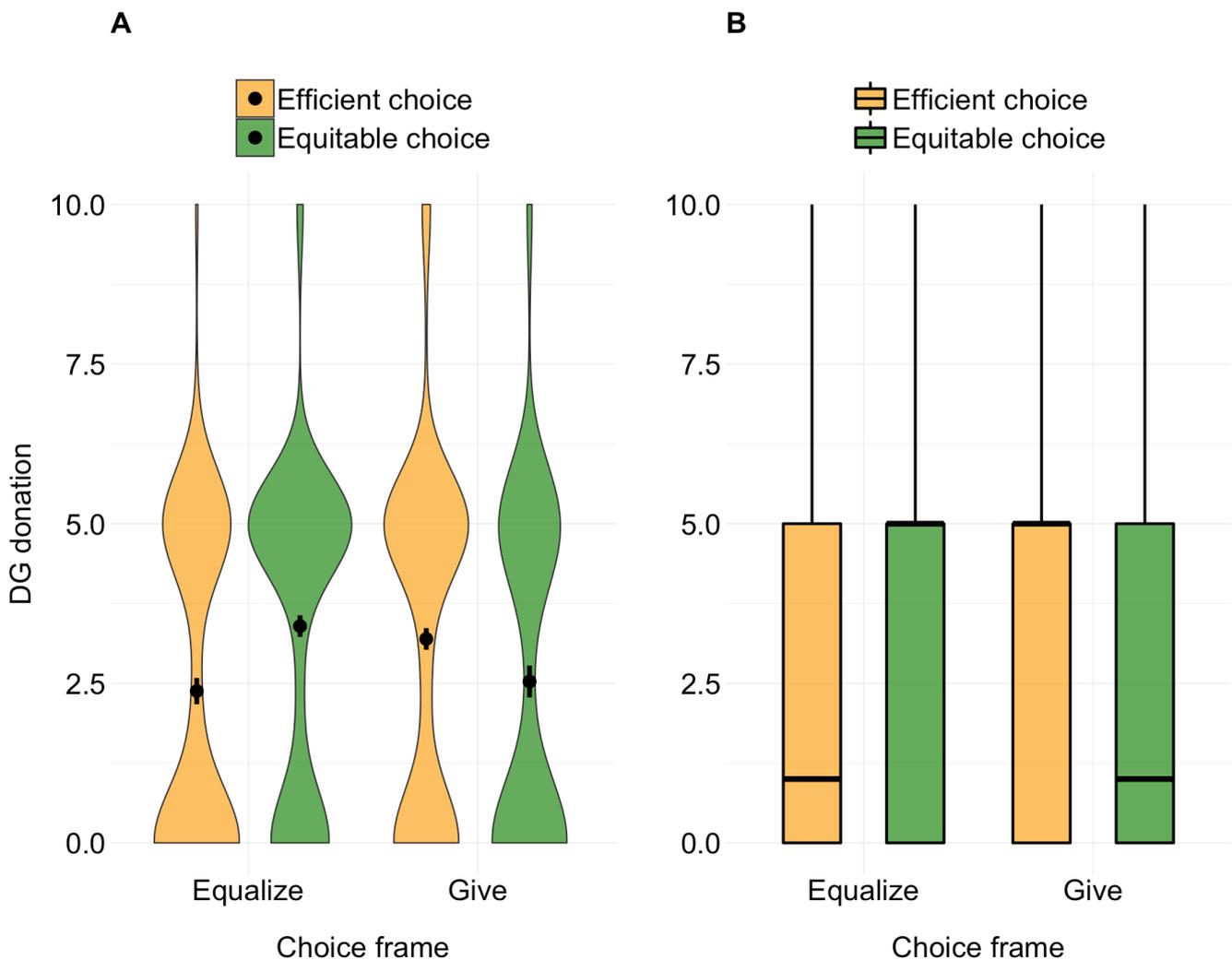

**Figure 2. Violin plots (A) and boxplots (B) of DG donations as a function of choice frame, and the choice the participants made, in the TOG. A**, *points denote the mean values and error bars are ± 1 SEM.* **B**, *bolded centre lines denote the median values.*

## Hypothesis 4

### Preregistered analyses

In our final preregistered analysis, we test whether the difference in DG donations between participants who made the 'moral' vs. 'non-moral' choice in the TOG is larger under the "do good" frame than under the "avoid bad" frame (Hypothesis 4). In line with our preregistered protocol, to simplify this analysis, we collapse across two variables: TOG choice frame [equalize, give], and the TOG choice the participant made [efficient, equitable]. This provides a new binary variable denoting whether the participant made the moral choice in the TOG [0=no, 1=yes], where the moral choice is simply defined as *either* the efficient option under the "give" frame, *or* the equitable option under the "equalize" frame. We then fit a linear regression model with two variables as predictors: moral choice [0=no, 1=yes], and moral frame [0=avoid bad, 1=do good]. As before, the DV is amount donated in the DG [0-10].

Inconsistent with Hypothesis 4, there is no statistically significant interaction between moral choice and moral frame on DG donations, b = -0.32, SE = 0.40, t = -0.79, p =.429. In other words, while participants who made the moral choice in the TOG tended to donate more in the DG than participants who made the non-moral choice (i.e., Hypothesis 3), this effect appeared largely independent of whether the participants made their TOG choice under the moral frame of "doing good" or "avoiding bad". Adjusting for age, gender, and education in the model did not meaningfully change this result, b = -0.33, SE = 0.40, t = -0.81, p =.418 (see SI for robustness checks).

## Discussion

Converging evidence suggests that prosociality in anonymous, one-shot interactions is not solely motivated by outcome-based social preferences, but that it is also motivated by what individuals perceive to be the morally right action (Bicchieri, 2005; DellaVigna et al., 2012; Eriksson et al, 2017; Kimbrough & Vostroktunov, 2016; Krupka & Weber, 2013); perhaps serving to maintain a positive moral self-image (Aquino & Reed, 2002; Dunning, 2007). Building on this work, recent experimental evidence advanced the hypothesis that a generalized morality preference drives prosocial behavior in anonymous, one-shot interactions like that in Dictator and Prisoner's Dilemma games (Capraro & Rand, 2018). This hypothesis rejects the classic view in behavioral economics that prosociality in these situations is driven by social preferences for equity and efficiency.

Here we identified prominent methodological confounds and open theoretical questions in the key experiments reported in Capraro and Rand (2018). In particular, their Trade-Off Game (TOG) design (i) confounds preferences for efficiency with self-interest, and (ii) preferences for action with preferences for morality. Moreover, the design fails to dissociate preferences to do "good" from preferences to avoid doing "bad". It is highly likely these issues affected the observed association between choice in the TOG, and prosociality in the DG/PD; the key evidence for the morality preference hypothesis. Likewise, the failure to decouple the preference to do "good" from that to avoid doing "bad" leaves the mechanism of the proposed morality preference unclear, and misses a key prediction from influential psychological theory; that "bad is stronger than good" (Baumeister et al., 2001; Rozin & Royzman, 2001; Vaish et al., 2008).

To address these issues, we designed and implemented an improved TOG/DG experiment – eliminating the confounds identified in the original experiments reported in Capraro and Rand (2018). In doing so, we replicated the key results in support of the morality preference hypothesis. We found that framing one or the other TOG choice as morally appropriate – by labelling the focal choice "fair" or "generous", or the counterpart choice "unfair" or "ungenerous" – strongly affected individuals' choices. Specifically, approximately 70% of individuals chose the efficient option when that choice was framed as morally appropriate, dropping to 40% when the equitable choice was framed as morally appropriate; a swing of 30%, and a reverse in the majority decision. More importantly, we found that individuals who chose the morally appropriate option in the TOG – be that the efficient option *or* the equitable option – were more prosocial in the preceding DG; donating more money to a stranger. This result was robust to various analytic specifications, and provides evidence that prosociality (in the DG) is driven by a preference for doing what is perceived to be morally right.

Our results lend experimental support to various alternatives to outcome-based preference models (e.g., Alger & Weibull, 2013; Brekke et al., 2003; Kimbrough & Vostroktunov, 2016; Krupka & Weber, 2013; Lazear et al., 2012; Levitt & List, 2017). In particular, these models assume that individuals have moral preferences that guide their prosocial decision-making in unilateral interactions; an assumption consistent with the current findings. A parallel class of models has sought to explain prosocial decision-making using an intention-based framework, according to which people are sensitive to others' intentions (Falk et al., 2008; McCabe et al., 2003; Rabin, 1993). These models have been useful in explaining prosociality observed in interactions with more than one active player. However, they are of limited use in the case of unilateral interactions – which are the focus of the current study – where beliefs about the intentions of others do not apply. Reputation-based models (e.g., Heck & Krueger, 2017; Jordan et al., 2016; Nowak & Sigmund, 2005) also appear of limited use for our results because choices were anonymous and one-shot.

Our findings are consistent with work in social psychology that suggests individuals are motivated to maintain a positive moral self-image (Dunning, 2007). Indeed, an open question at the intersection of this research is whether individuals who assign greater value to a moral self-image (Aquino & Reed, 2002) show stronger choice framing effects in the TOG, and/or a stronger association between TOG choice and prosociality in the DG/PD. That said, we note that it is unlikely individuals are *solely* motivated by what they perceive to be the right thing, especially across different choice contexts. In line with Capraro and Rand (2018), we do expect that these results extend to the PD. There is some recent evidence, however, that behavior in other economic games is driven by outcome-based preferences, and *not* by a general morality preference (Capraro, 2018). An interesting avenue for future work is thus to further explore the boundary conditions of the morality preference account. One avenue might be situations in which people have to trade-off conflicting moral principles, like in the case of "altruistic" lying; that is, lying to benefit others (Biziou-van-Pol et al., 2015; Erat & Gneezy, 2012). An open empirical question is whether and how preferences for morality – as revealed by choice in the TOG – predict moral trade-offs of this kind.

We found that a moral frame emphasizing "good" affected TOG choice as strongly as a moral frame emphasizing "bad". For example, the proportion of individuals switching from the efficient choice to the equitable choice was essentially identical (approx. 30%) whether the latter choice was labelled "fair" or the former choice "unfair". We thus found little evidence that "bad" (framing) was stronger than "good" (framing) in prosocial choice (cf. Baumeister et al., 2001).

An explanation for this discrepancy is found in recent work that has criticized the dominance of "bad" over "good" (termed the *negativity bias*) on theoretical and empirical grounds (Corns, 2018). A particular criticism concerns the credible alternative explanations for much of the evidence base. For example, asymmetries in perception that are taken to support the stronger effect of negatively-valenced stimuli may instead be explained by differences in the informativeness of negative vs. positive stimuli (Corns, 2018). In research on impression formation, as a case in point, the greater weight assigned to immoral (vs. moral) traits may be explained by the fact that immoral traits tend to be more *informative* of others' character (Kellermann, 1984). In the case of *self*-perception, in contrast, it is likely this informativeness bias does not hold; for example, because individuals can introspect on their full "moral history", unlike when they are judging other people. Absent a difference in informativeness, immoral and moral traits may yield similar impact on judgment and behavior.

This provides a plausible explanation for our results, and is consistent with a recent critique of the negativity bias hypothesis (Corns, 2018). This explanation also clarifies why the observed association between choosing the morally-framed option in the TOG, and prosociality in the DG, was similar under the "good" vs. "bad" moral frame (i.e., rejection of H4). Specifically, because the frames were *equally* motivating, individuals choosing the morally-framed option in either frame were *equally* liable to donate more in the DG. It is important to highlight that our "bad" frame label was *symmetric* vis-à-vis our "good" frame label (e.g., "unfair" vs. "fair", respectively). Had we used frames with stronger, but asymmetric labels – such as "steal" in place of "unfair" – we may have elicited an asymmetric moral frame effect in line with the negativity bias hypothesis. As noted above, however, in that case it may have been unclear whether the moral frame asymmetry was due to the valence of the frames per se (i.e., bad vs. good), or *other* differences such as asymmetric strength or salience of the labels (cf. Corns, 2018).

Finally, our results add to a growing body of research on framing effects in prosocial choice. Prior work finds that situational labels sometimes impact prosocial decision-making (e.g., Capraro & Vanzo, 2018; Kay & Ross, 2003; Krupka & Weber, 2013; Larrick & Blount, 1997; Liberman et al., 2004), but not always (Dreber et al., 2013; Goerg et al., 2017). This suggests the effect of labels may depend on the specific interaction or game type. Our findings replicate recent work showing that labels are highly effective in the Trade-Off Game (Capraro, 2018; Capraro & Rand, 2018).

In summary, recent experimental work advanced the hypothesis that prosociality in anonymous, one-shot interactions is not driven by outcome-based social preferences for equity or efficiency per se, but by a generalized morality preference for "doing the right thing" (Capraro & Rand, 2018). We identified prominent methodological confounds and open theoretical questions in this work, and, consequently, conducted a refined and extended test of the morality preference hypothesis. Consistent with this hypothesis, our findings indicate that prosociality in the anonymous, one-shot Dictator Game is driven by preferences for doing the morally right thing. Furthermore, consistent with a recent critique of the negativity bias hypothesis, our results suggest the preference to do "good" was *as* potent as the preference to avoid doing "bad" in this case.